\documentclass[epj]{svjour}
\usepackage{graphics}
\begin{document}
\title{Evidence for Conservatism in LHC SUSY Searches}
\author{Benjamin Nachman\inst{1} \and  Tom Rudelius\inst{2}% etc
}                     % Do not remove
\institute{SLAC, Stanford University \and Department of Physics, Harvard University}
\date{Received: 18 November 2012 / Revised: 1 December 2012}
\abstract{
The standard in the high-energy physics community for claiming discovery of new physics is a $5\sigma$ excess in the observed signal over the estimated background.  While a $3\sigma$ excess is not enough to claim discovery, it is certainly enough to pique the interest of both experimentalists and theorists.  However, with a large number of searches performed by both the ATLAS and CMS collaborations at the LHC, one expects a nonzero number of multi-$\sigma$ results simply due to statistical fluctuations in the no-signal scenario.  Our analysis examines the distribution of p-values for CMS and ATLAS supersymmetry (SUSY) searches using the full 2011 data set to determine if the collaborations are being overly conservative in their analyses.  We find that there is a statistically significant difference between the expected and observed distributions of p-values and suggest that the most probable cause is over-conservatism in the estimation of uncertainties.
%
%\PACS{
 %     {PACS-key}{discribing text of that key}   \and
   %   {PACS-key}{discribing text of that key}
     %} % end of PACS codes
} %end of abstract
\maketitle

\section{Introduction}

The Large Hadron Collider (LHC) is currently the world's most energetic particle accelerator, colliding beams of protons at a center of mass energy of 8 TeV.  In addition to probing the mechanism for electroweak symmetry breaking (discovering the Higgs Boson), one of the LHC's purposes is to probe the current model of particle physics and possibly find new physics.  There are two multi-purpose detectors that observe particle collisions at the LHC: A Toroidal LHC Apparatus (ATLAS) and the Compact Muon Solenoid (CMS).  These detectors measure the properties of decay products resulting from the proton collisions.  The particle content is a stochastic process.  The physics which is most interesting occurs at a small rate and so to generate a large expected number of events one needs to collect a large quantity of data.  This is precisely what both collaborations have done since the LHC turned on in 2010 - collecting over $10/\mathrm{fb}$ of data.  In this paper, we perform a statistical analysis on ATLAS and CMS results which seek to discover new physics at the LHC.  

The most common form of new physics that is sought after at the LHC is Supersymmetry (SUSY).  SUSY is a symmetry linking each boson to a fermion, known as its `superpartner.'  Originally, the supersymmetric standard model was proposed to solve the hierarchy problem.  Since then, it has been invoked to account for dark matter and to stabilize bosonic string theories under the new title of superstring theory.  Thus, SUSY can be applied to solve many of the open problems in high-energy physics.  However, if SUSY is truly realized in nature as a solution to the hierarchy problem, then superpartners should start to appear near the energies currently being probed by the LHC.  Therefore, both ATLAS and CMS have formed large working groups to search for signs of these new particles in the collision data.  No evidence for SUSY has yet been found.

Generically, there are three steps of a SUSY search at the LHC.  The first step is to place restrictions on the data in order to isolate potentially interesting events.  This is done by identifying properties of events which new physics might have, but known physics will not possess.  This process is often driven by using Monte Carlo data that simulates new physics scenarios.  Once criteria are chosen, the next step is to estimate the number of known physics events which will pass this criteria.  There are many techniques for this estimation, some which depend heavily on Monte Carlo data and some which are `data-driven'.  The final step is to estimate the uncertainty in the estimation of the number of predicted known physics events.   ATLAS and CMS papers report the number of observed events which pass the selection criteria, the expected number of events and a corresponding uncertainty.  The distribution of these uncertainties is never reported in publications, so they must be treated as Gaussian by particle physicists reading the results.  The Gaussian uncertainty is a good approximation, but there is a hard bound at zero so that a negative number of events is not considered.

If the number of observed events is much larger than the expected number of events, there is evidence to claim discovery.  The number of events observed should follow a Poisson distribution whose mean follows a Gaussian distribution of appropriate mean and variance.  By convolving the Poisson and Gaussian distributions, we calculate a p-value for each point in our data set.  We analyze this distribution of p-values, finding that the observed distribution differs significantly from the expected one.

\section{Constructing the Data Set}

The data are chosen out of the set of all published\footnote{We also consider results approved by the collaborations for journal submission even if they have not yet been published.} SUSY searches from ATLAS and CMS on the full 2011 LHC data set.  The 2011 data set was chosen because at the time of writing, no analyses have been published on the 2012 data set, since it takes many months to analyze and publish results.  There are many more public SUSY searches by both collaborations in the form of Conference Notes (ATLAS) and Public Analysis Summaries (CMS), but these results are intended to be preliminary.  Therefore, we consider only the final results that appear in journals.  There are 14 such papers, 7 each from the two collaborations.  Each of these papers describes a search in which there were several {\it signal regions} that essentially constitute multiple searches.  The difficulty in conducting an analysis of all the signal regions over all the papers is to correctly consider the correlations between studies.   Since publications often do not give every detail of the analyses, we use the information given to reduce the full set of analyses to a maximal set of uncorrelated ones.  There will necessarily be some arbitrary choices made in this process, but in the end we find the same results for analysis on the maximal uncorrelated set as we do on the full data set.

The general strategy in disentangling the signal regions is to first look at which objects were studied in a given analysis and then to compare the requirements on those objects.  The objects are [charged] leptons, jets and missing energy.   The leptons are electrons, muons and taus.  Analyses can consider any number of leptons and various other requirements such as charge, momentum, invariant mass, etc.   Jets are the result of modeling the hadronization and fragmentation of quarks and gluons inside the detector.  These objects are constructed by grouping many particles that were close enough in space.  The sum of the particles is then treated as one object, with a momentum and an energy.  Selection criteria on jets are similar to leptons with properties like multiplicity, momentum and invariant mass.  Other criteria include flavor (distinguishing $b$ quark jets from lighter quark or gluon jets) and isolation from other objects.  The final object that we need to partition the analyses is the missing momentum, often called MET, $E_T^{miss}$ or $ {E}_T\hspace{-4.5mm}\slash$\hspace{3mm}.  This object is the vector in the plane transverse to the beam pipe which is opposite in direction to the sum of the transverse vectors of everything else observed in the detector.  It is the transverse momentum necessary to conserve momentum in the plane transverse to the beam pipe.  Sources of MET in an event include neutrinos and also new physics particles, which do not interact often with normal matter.  Furthermore, a nonzero MET can result from the incomplete calorimeter and tracker coverage in both the transverse and azimuthal directions as well as mismeasurement of the transverse momentum of jets and other objects.  Criteria on the MET include its magnitude, direction relative to the other objects, and invariant mass with the leptons.  This last variable is often computed using transverse quantities and is given the name {\it transverse mass} and denoted $m_T$.

The process for constructing the maximally independent data set involves a top-down partitioning of the full set of signal regions over all analyses.  The most obvious partition is between ATLAS and CMS analyses.  Except for some shared theoretical uncertainties in background estimates and uncertainty in deriving the LHC luminosity measurement, analyses are uncorrelated between the two collaborations.  The luminosity uncertainty should be minimal for background estimates that derive from or calibrate to the data.  Within a given collaboration there will be some correlation in the uncertainty from detector effects.  We will ignore such correlations in the following analysis.

A next clear partition is based on the multiplicity of leptons.  Analyses which require leptons are independent from those which veto the presence of a lepton.  Given the same criteria on the reconstruction on the leptons, this is an exact statement for the data.  However, there are slight correlations in the background estimates, since fake and lost leptons are backgrounds for various analyses.  However, this potential overlap is quite small and so is ignored in selecting independent signal regions.   This same logic applies to jet multiplicity.  Given the same jet reconstruction, analyses which require exactly two jets are declared independent from analyses which require three or more jets.  Once again, this is affected by energy cuts on the jets and also on the region of the detector used to reconstruct these objects.  Such details are mostly ignored in choosing the signal regions.  

The only two analyses which allow for more than two leptons are~\cite{ATLASmultilep} and ~\cite{CMSmultilep}.  The first of these analyses has only three signal regions, all of which are non-overlapping based on requirements on the invariant mass of set of pairs of the leptons.   The CMS paper performs two analyses by partitioning the data in two ways.  Since these two analysis use essentially the same events,  we choose one of them to include in our minimal independent dataset.  We arbitrarily choose the first analysis, which uses $E_T^{miss}$ a variable called $H_T$ and the invariant mass of subsets of the leptons to partition the data into $52$ signal regions.  Between these two multi-lepton searches, we therefore consider $55$ signal regions.  

While ATLAS has performed one analysis requiring exactly two leptons~\cite{ATLAS}, CMS has four published papers with dilepton searches~\cite{CMS2leps1}~\cite{CMSZleptons}~\cite{CMSSSleptons}~\cite{CMSSSleptonsB}.   Both~\cite{CMSSSleptons} and~\cite{CMSSSleptonsB} require same sign dileptons and so to minimize correlations we pick all 27 of the signal regions in~\cite{CMSSSleptons} for our minimal set.   The remaining CMS analyses require oppositely charged leptons, but are distinguished in that~\cite{CMS2leps1} vetos leptons with an invariant mass near the mass of the Z boson and~\cite{CMSZleptons} requires such a mass.  Some of the regions in~\cite{CMSZleptons} and~\cite{CMS2leps1} are correlated.  We therefore use the following procedure to de-correlate them.  Suppose that we have two signal regions defined by $X>x_1$ and $X>x_2$, $x_1<x_2$, {\it ceteris paribus}.  If region $i$ has an expected background of $b_i\pm \sigma_i$ and $d_i$ observed counts then we define two new regions $x_1<X<x_2$ and $X>x_2$ with $b_1-b_2\pm\sqrt{\sigma_1^2-\sigma_2^2}$ as the estimated background in the first region and $b_2\pm \sigma_2$ estimated in the second region.  The number of observed counts in the first of these new regions is $d_1-d_2$ and in the second region this number is $d_2$.  With this procedure we are able to de-correlate all 10 regions in~\cite{CMSZleptons} and all 8 signal regions in~\cite{CMS2leps1} bringing our total from CMS dilepton searches to 45 signal regions.  We then add the 6 uncorrelated dilepton signal regions from ATLAS~\cite{ATLAS} to bring the total to 51 signal regions from dileptons.

Of the 2012 published papers, the only single lepton searches were performed with the ATLAS detector~\cite{ATLAS1lep}~\cite{ATLAS}.  Since these analysis are correlated, we take most of the signal regions from~\cite{ATLAS1lep} and then add in the soft lepton signal region from~\cite{ATLAS}.  The de-correlation procedure described in the previous paragraph is applied to the five signal regions in~\cite{ATLAS1lep}.  These two ATLAS analyses thus give $7$ total signal regions to the maximal uncorrelated set.

The rest of the analyses veto the presence of a lepton and so we need another criterion to partition the signal regions.  First, we consider the two zero lepton searches in CMS~\cite{CMS0leptons}~\cite{CMS0leptonsMT2}.  Since it is hard to discern the exact correlation between events, we simply take all 14 uncorrelated signal regions from~\cite{CMS0leptons}.  Next, we add to our list from the zero lepton ATLAS searches~\cite{ATLAS3btags}~\cite{ATLAStopcut}~\cite{ATLASglus}~\cite{ATLASjets}.  Many of the signal regions in these searches have very similar criteria.  Therefore, we pick from this set the high jet multiplicity signal regions from~\cite{ATLASjets} and low jet multiplicity signal regions from~\cite{ATLASglus}.  There remains some residual correlation because the latter paper does not veto high multiplicity jets.  Therefore, we take regions separated by two in jet multiplicity threshold to reduce correlation.  In particular, we only take signal regions A and C from~\cite{ATLASglus}.  This is the only set of signal regions which could have residual overlap.  The overlap is expected to be small and there are only a few signal regions in question.  The de-correlation procedure is applied to both sets of regions selected.  This brings the total number of signal regions in our maximally independent data set to 137.

\section{Results}

The formula for the p-value associated with a SUSY search signal region with observed number of counts $n$, expected number of counts $\mu$, and uncertainty $\sigma$ is given by the convolution,
\begin{equation}\label{convolution}
\mbox{p-value} = \int_{0}^{+\infty}{\phi(\lambda|\mu,\sigma)P_n(\lambda)d\lambda}.
\end{equation}
Here, $\phi(\lambda|\mu,\sigma)$ is the probability density function (p.d.f.) of the normal distribution with mean $\mu$, standard deviation $\sigma$,
\begin{equation}
\phi(\lambda|\mu,\sigma) = \frac{1}{\sigma\sqrt{2\pi}}e^{-(\lambda-\mu)^2/2\sigma^2},
\end{equation}
and $P_n(\lambda)$ is the probability of observing $n$ or more counts given a Poisson distribution with parameter $\lambda$,
\begin{equation}
P_n(\lambda) = \displaystyle\sum_{k=n}^\infty{\frac{e^{-\lambda} \lambda^k}{k!}} = 1-\displaystyle\sum_{k=0}^{n-1}{\frac{e^{-\lambda} \lambda^k}{k!}}.
\end{equation}
By this method, we associate a p-value with each trial.  For a continuous probability distribution, it is well known that p-values follow a uniform distribution on the interval $[0,1]$ under the null hypothesis~\cite{Hartung}.  This is intuitively clear because the p-value $x$ represents the probability of observing a result less than $x$, which is precisely the linear cumulative distribution function associated with the uniform distribution.  However, for a discrete probability distribution such as the Poisson distribution used here, only discretely many p-values between $0$ and $1$ are possible.  This leads to non-uniformity in the distribution of p-values, especially for trials with small $\mu$.  To account for this, we must first compute the expected distribution of p-values under the null hypothesis and compare this with the observed distribution of p-values.

\begin{figure}[H]
\begin{center}
\resizebox{0.45\textwidth}{!}{
\includegraphics{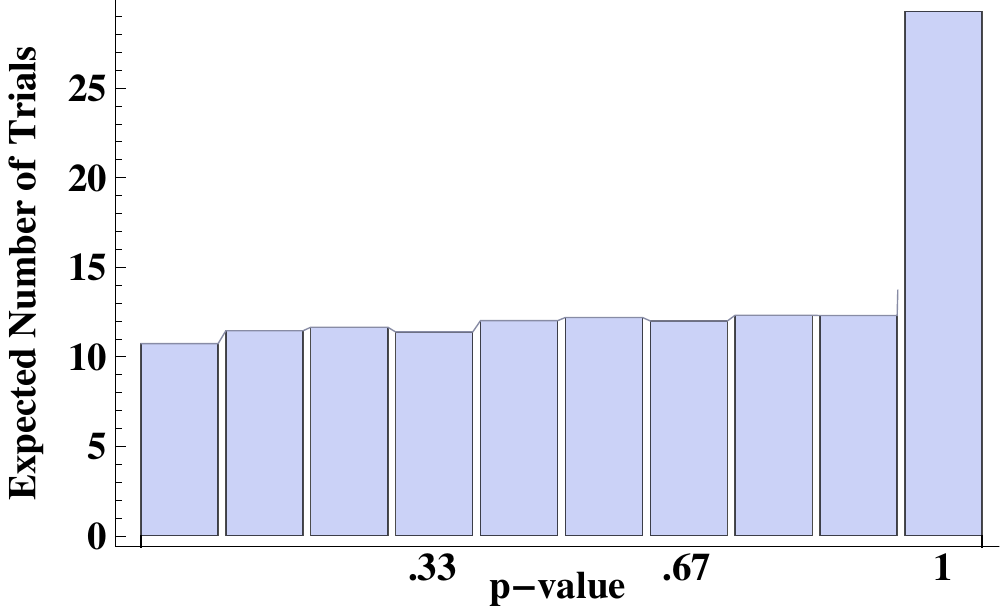}
}
\caption{Expected distribution of p-values.  The distribution of p-values expected under the null hypothesis is largely uniform, except for the spike at 1 due to trials with $\mu < 1$. }
\label{expected}
\end{center}
\end{figure}

To compute the expected distribution of p-values, we calculated the expected distribution of p-values for each trial and averaged over all trials.  For a given trial with expected mean $\mu$ and uncertainty $\sigma$, we calculated the probability that the p-value of the trial would fall within the range $(\frac{i}{10},\frac{i+1}{10}],i=0,1,...,9$ according to,
\begin{equation}
\Pr\left(\frac{i}{10}<\mbox{p-value}\leq\frac{i+1}{10}\right) = \frac{1}{N}\int_0^\infty{d\lambda f_i(\lambda)\phi(\lambda|\mu,\sigma)}.
\end{equation}
The normalization constant $N$ sets the sum $\sum_i{f_i}=1$ and is necessary to account for the fact that the parameter $\lambda$ of the Poisson distribution cannot be negative.  Once again, $\phi(\lambda|\mu,\sigma)$ is the p.d.f. of the normal distribution $N(\mu,\sigma^2)$ and $f_i(\lambda)$ is given by,
\begin{equation}
f_i(\lambda)=\displaystyle\sum_{m=0}^\infty{\left[\Pr(X=m)\cdot \Pr\left(\frac{i}{10}<\Pr\left(X\geq m\right) \leq\frac{i+1}{10}\right)\right]},
\end{equation}
where $X \sim \mbox{Poisson}(\lambda)$.  Note that the second term in the product of the summand is either $0$ or $1$ depending on the value of $\Pr\left(X\geq m\right)$.  For the sake of calculation, we approximated the distribution of p-values for trials with $\mu > 10$ as uniform on $[0,1]$.  Numerical analysis showed this to be a very good approximation.  After averaging over all trials, we computed the expected distribution of p-values under the null hypothesis shown in Figure \ref{expected}.  The spike at 1 is due to the fact that many trials had $\mu < 1$ (scaled MC predictions allow for fractional events $0<\mu<1$), and it is very likely that such a trial would yield $n=0$, corresponding to a p-value of 1.

\begin{figure}[H]
\begin{center}
\resizebox{0.6\textwidth}{!}{
\includegraphics{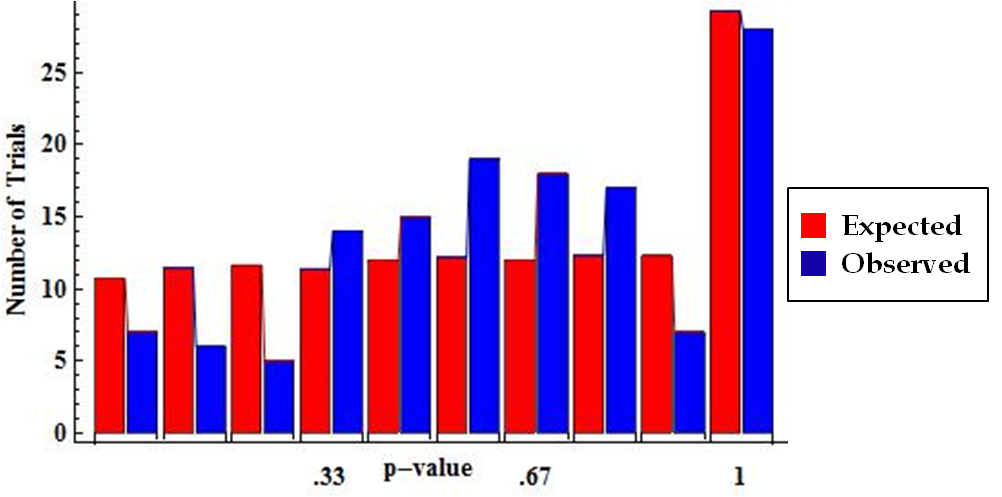}
}
\caption{Expected vs. observed distribution of p-values.  The distribution shows a deficiency of p-values in the tails and in the lower bins, indicating conservatism in the reported background and uncertainty estimates.}
\label{observed}
\end{center}
\end{figure}

The observed distribution of p-values is contrasted with the expected distribution in Figure \ref{observed}.  Figure \ref{observed} graphically confirms our suspicions that the reported background and uncertainty estimates may be too conservative.  The deficiency of small p-values shows that the true background values may be smaller than the reported values, and the excess of p-values in the center of the distribution shows that the true uncertainty values may be smaller than the reported values.

Table \ref{results} quantifies the information in Figure \ref{observed}.  From a $\chi^2$ goodness-of-fit test, we conclude that the expected and observed distributions are significantly different at a significance level of $0.05$.  At this point, we perform a post-hoc analysis to determine the most likely cause for this discrepancy.  Ostensibly, an over-estimation of the mean background values would result in a deficiency of low p-values, while an over-estimation of the background uncertainties would result in a deficiency of p-values in the tails of the distribution.  Thus, to test the mean background estimates, we compare the observed number of trials with p-value less than $.2$, $.3$, $.4$, and $.5$ to the expected number for each of these categories under the null hypothesis $H_0$, finding marginally significant results.  To test the uncertainty estimates, we compare the observed number of trials with p-value less than $.2$ or greater than $.8$, finding more compelling evidence for conservatism.  These conclusions hold regardless of whether we use the reduced data set (after correlated data points are removed) or the complete data set (before correlated data points are removed).  Table \ref{results} gives the distance $t$ in standard deviations from the expected value as well as the probability that a distance $t$ or greater would be observed under $H_0$.

\begin{table}[H]
\begin{center}
\begin{tabular}{||c|c|c|c||} \hline
Quantity & Dist. under $H_0$ ($T$)& Test statistic ($t$)&P$(|T|>|t|)$\\ \hline
Trials with $p<0.1$&N(0,1)&-1.186&0.2355 \\ \hline
Trials with $p<0.2$&N(0,1)&-2.130&0.0332 \\ \hline
Trials with $p<0.3$&N(0,1)&-3.138&0.0017 \\ \hline
Trials with $p<0.4$&N(0,1)&-2.403&0.0162 \\ \hline
Trials with $p<0.5$&N(0,1)&-1.774&0.0761 \\ \hline
Trials with $0.2<p<0.8$&N(0,1)&2.828&0.0047 \\ \hline
Expected vs. observed distribution&$\chi_9^2$&19.959&0.018 \\ \hline

\end{tabular}
\caption{Results of statistical hypothesis tests.  The first $6$ rows report a post-hoc analysis comparing the number of trials with p-value less than $0.1, 0.2, 0.3, 0.4$, and $0.5$ and the number of trials with $0.2<p<0.8$ to the numbers that would be expected under the null hypothesis $H_0$.  The last row is a $\chi^2$ goodness-of-fit test comparing the entire expected and observed distributions.  For all the tests, the distance $t$ in standard deviations from the expected value is reported, as is the probability that a distance $t$ or greater would be observed under $H_0$.}
\label{results}
\end{center}
\end{table}

\section{Conclusion}

We have performed a statistical analysis of p-values from 2011 SUSY searches at the LHC to investigate the accuracy of predictions under the null hypothesis of no SUSY.  We find that the observed and expected distributions of p-values are significantly different at a level of $0.05$.  This indicates that the hypothesis of normally-distributed backgrounds with the reported background and uncertainty estimates should be rejected.  There are two possible explanations for this result: either the background uncertainties are not normally distributed or else the background uncertainties are normally distributed, but the reported mean background values and/or uncertainty values are inaccurate.  In the former case, our post-hoc analysis of the observed distribution suggests that the true distribution would have to be lighter-tailed that the assumed Gaussian distribution to explain the discrepancy of p-values in the tails of the observed distribution.  While distributions with lighter tails than the Gaussian distribution certainly exist, they are relatively rare.

The alternative conclusion, which we find more plausible, is that the background uncertainties are indeed (approximately) normally distributed, but the reported background values and/or uncertainties are inaccurate.  Again, our analysis suggests that uncertainty overestimation is the most probable explanation for the observed data.  However, due to the post-hoc nature of this particular investigation, we express caution in presenting our theory of the cause of the discrepancy between the expected and observed distributions of p-values.

We encourage future publications of SUSY search results to report the (approximate) distributions of background uncertainties whenever possible.

\section{Acknowledgements}

We would like to thank Till Eifert, Matt Reece and Kyle Cranmer for the helpful discussions.  This material is based upon work supported by the National Science Foundation under Grant No. DGE-1144152.

%
% For  figures use
%\begin{figure*}
%\vspace*{5cm}       % Give the correct figure height in cm
%\includegraphics{leer.eps}
%\end{figure*}
% or  this
%\begin{figure}
%\centering
%\resizebox{0.75\textwidth}{!}{%
%  \includegraphics{leer.eps}
%\vspace{5cm}       % Give the correct figure height in cm
%\end{figure}
%\begin{table}
%\centering
%\begin{tabular}{lll}
%\hline\noalign{\smallskip}
%\end{tabular}
% Or use
%\vspace*{5cm}  % with the correct table height
%\end{table}

%
% BibTeX users please use
% \bibliographystyle{}
% \bibliography{}

\end{document}